\documentclass[11pt,twoside,a4paper]{article}
\usepackage{graphicx}
\usepackage{hyperref}
\usepackage{amsmath}
\usepackage{xcolor}
\usepackage[normalem]{ulem}

\title{Characterisation of Inhomogeneities in Ti:LiNbO$_3$ waveguides.}
\author{Matteo Santandrea$^{1,*}$, Michael Stefszky$^1$, Gana\"el Roeland$^2$, Christine Silberhorn$^1$}
\date{$^1$ Integrated Quantum Optics, Paderborn University, Warburgerstr. 100, Paderborn, Germany\\$^2$ Physics department, ENS-PSL Research University; 45 Rue d'Ulm, 75005 Paris, France\\
$^*$ Corresponding author: matteo.santandrea@upb.de}

\begin{document}
\maketitle
\section{Abstract}
Nonlinear processes in integrated, guiding systems are fundamental for both classical and quantum experiments. Integrated components allow for compact, modular and stable light-processing systems and as such their use in real-world systems continues to expand. 
In order to use these devices in the most demanding applications, where efficiency and/or spectral performance are critical, it is important that the devices are fully optimized. In order to achieve these optimizations, it is first necessary to gain a thorough understanding of current fabrication limits and their impact on the devices final performance.
In this paper we investigate the current fabrication limits of titanium in-diffused lithium niobate waveguides produced using a masked photolithographic method. 
By dicing a long ($\sim$8cm) sample into smaller pieces and recording the resulting phasematching spectra, the fabrication error present in the UV photolithographic process is characterized.
The retrieved imperfections fit well with theoretical expectations and from the measured imperfection profile it is shown that one can directly reconstruct the original distorted phasematching spectrum.
Therefore, our measurements directly quantify the intrinsic limitations of the current standard UV photolitographic technique for the realization of integrated waveguides in lithium niobate.

\section{Introduction}
Guiding systems fabricated in nonlinear materials are employed in a wide variety of contexts for both classical and quantum applications. In comparison to bulk systems, guiding systems provide simple integration into fiber networks and provide a tighter spatial mode confinement, which generally strengthens the nonlinear interaction. Such systems are already in use in a myriad of classical applications, for example in second harmonic and sum frequency generation (SHG/SFG) to efficiently produce light in spectral ranges otherwise inaccessible \cite{Lim1989, Eger1994, Sun2012}. More recently these systems are also finding applications in quantum systems, for example in generation and manipulation of quantum states \cite{Tanzilli2001,Tanzilli2005, Rutz2017, Maring2017, Stefszky2017}. 

While waveguiding systems offer many advantages, these systems also generally exhibit higher losses and are much more sensitive to device imperfections, when compared to their bulk counterparts. It is known that small device imperfections can dramatically reduce device performance \cite{Helmfrid1991, Helmfrid1992, Pelc2010, Cui2012, Phillips2013, Chang2014, FrancisJones2016, Santandrea2019}. 
For this reason, it is critical to assess the limits of the current fabrication technology to identify the classes of nonlinear processes physically achievable and to devise strategies to overcome such limitations. 

Despite the importance of these investigations, very little experimental work has been undertaken \cite{Chang2014, FrancisJones2016}.
In \cite{Chang2014}, both the amplitude and the phase of a phasematched process in reverse proton exchanged lithium niobate (RPE-LN) waveguides was characterized, allowing the complete determination of the inhomogeneity profile of their waveguides. However, such a scheme is not always possible as it relies on a chirped broadband conversion process to map spatial inhomogeneities to the amplitude and phase of the second harmonic field, which can be characterized using frequency resolved optical gating (FROG).
In \cite{FrancisJones2016}, a destructive approach was used to reconstruct the variation of the fabrication parameters of a photonic crystal fibre (PCF).
In this method, the sample under analysis is diced into smaller sections and the phasematching of each section is used to infer the local properties of the system.

In this paper, the dicing technique is used to retrieve the fabrication errors of titanium indiffused lithium niobate (Ti:LN) waveguides.
An 83mm-long sample is diced down into $\sim$10mm-long pieces, whose individual phasematching, as well as the phasematching profile of intermediate lengths, are characterized. 
The shift of the phasematching spectrum along the waveguide is mapped and used to retrieve the phasematching variation.
Via numerical modelling, we are able to relate the measured variation to fabrication parameter errors and from this show that the estimated errors agree with previous predictions \cite{Santandrea2019}.
Finally, from the measured waveguide inhomogeneities, we are able to reconstruct the original phasematching of the sample.

\section{Experiment}
The system under investigation is a set of seven 83mm-long nonlinear waveguides fabricated by in-diffusing Ti ions in a z-cut LiNbO$_3$ crystal. 
The waveguides are designed to be single-mode in the telecom C band ($7\mu$m in width) and are periodically poled with a period $\Lambda$=16.8$\mu$m. 
This allows a type 0, $ee\rightarrow e$ degenerate SHG process pumped at 1528.4nm at room temperature.

The nonlinear process is completely determined by the phase mismatch $\Delta\beta$ of the involved light fields
\begin{equation}
\Delta\beta(z, \lambda) = 2\pi\left(\frac{n_e(z, \lambda/2)}{\lambda/2} -2\frac{n_e(z, \lambda)}{\lambda}  -\frac{1}{\Lambda} \right),\label{eq:deltabeta}
\end{equation}
where $n_e(z,\lambda)$ is the extraordinary refractive index of LiNbO$_3$ at the position $z$ and at the pump wavelength $\lambda$.
We explicitly consider the variation of the refractive index along the propagation axis $z$ to include the effect of fabrication imperfections, such as inhomogeneities in the waveguide width, depth, operating temperature, poling pattern or a combination thereof. 

In the case of small refractive index variations, one can approximate variations in the momentum mismatch to be wavelength independent, i.e. $\Delta\beta(z, \lambda) \approx \Delta\beta_0(\lambda) + \delta\beta(z)$. In this case, the output intensity spectrum of the SH process is given by \cite{Helmfrid1992, Chang2014, Santandrea2019}
\begin{equation}
I(\lambda) \propto \left|\int_0^L \mathrm{e}^{-i\Delta\beta_0(\lambda) z}\mathrm{e}^{-i\int_0^z \delta\beta(\xi)\mathrm{d}\xi}\mathrm{d}z \right|^2 \label{eq:pm_integral}
\end{equation}
and thus depends on the specific $\delta\beta(z)$ profile. 
However, if the waveguide is sufficiently short, the impact of the variations $\delta\beta(z)$ are not strong enough to appreciably distort the phasematching spectrum, as shown in \cite{Santandrea2019}. They are still expected, however, to affect positioning of the centre of the spectrum.
Previous work in this waveguide system has shown that we expect to see nearly ideal phasematching spectra for waveguides shorter than 1cm in length \cite{Santandrea2019}.
Therefore, it should be possible to retrieve the variation of $\delta\beta(z)$ along the sample by dicing a long sample down to $\sim$1cm-long pieces and monitoring the shift in the position (in wavelength) of the phasematching spectrum.

\begin{figure}[bt]
\centering
\includegraphics[width = 0.7\textwidth]{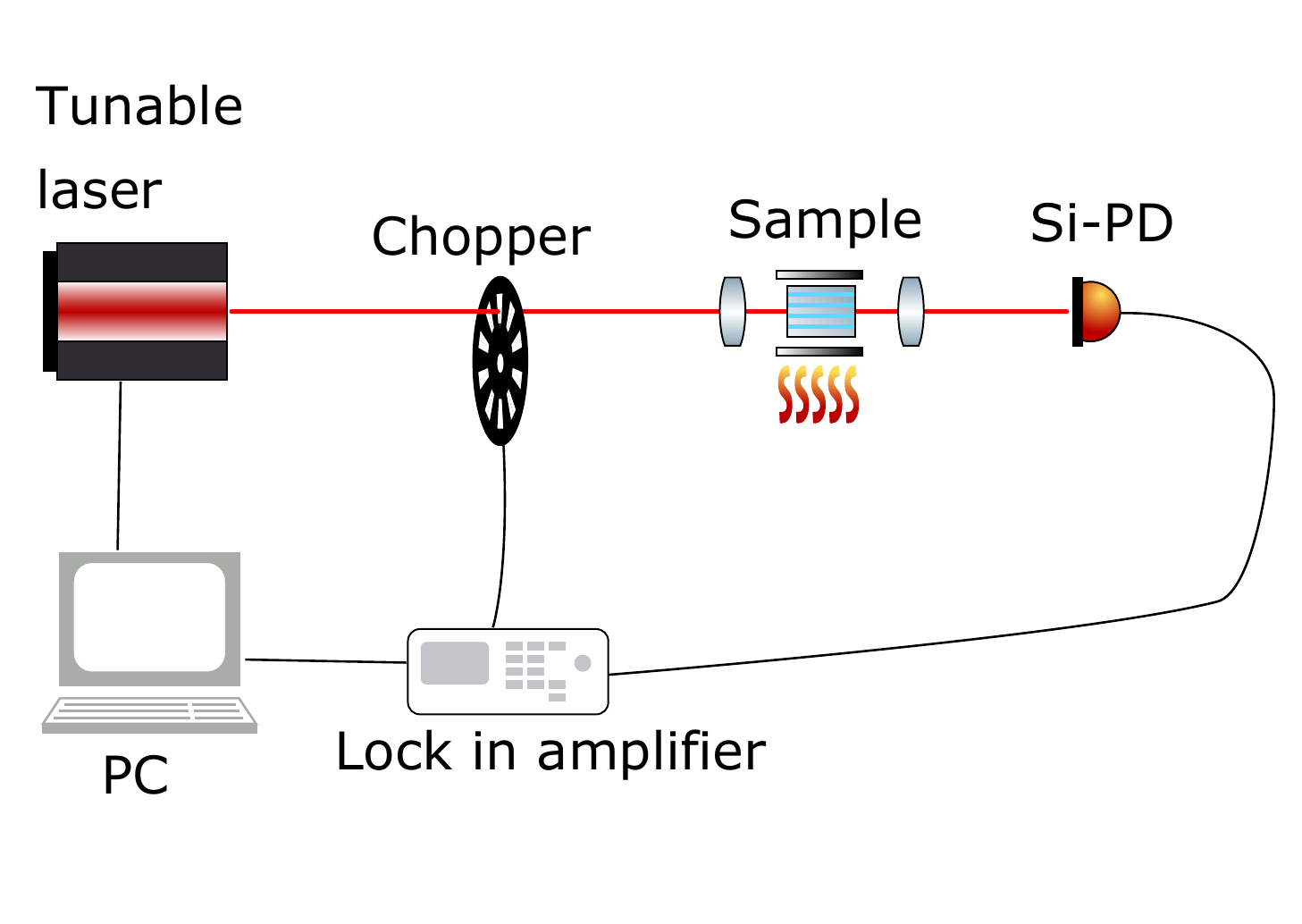}
\caption{Measurement setup. Full description in text.}
\label{img:setup}
\end{figure}

At first, the phasematching spectra of the initial 83cm-long waveguides are measured in the setup illustrated in Figure \ref{img:setup}. The sample is first stabilised in temperature to around 25$\pm$0.1$^\circ$C as unwanted temperature shifts are indistinguishable from true waveguide imperfections, since both will act to shift the centre of the phasematching profile. The waveguides are then pumped with approximately 3mW from a tunable 1550nm laser (\textit{EXFO Tunics}) and the wavelength of the pump is scanned in 1pm steps. The generated SH field is then detected using a Si-PIN photodiode. To increase the signal-to-noise ratio of the measurement the pump field is run through a chopping blade, and the photocurrent from the photodiode is fed into a lock-in amplifier (\textit{Ametek Scientific Instrument 7265}).

\begin{figure}[bt]
\centering
\includegraphics[width = 0.7\textwidth]{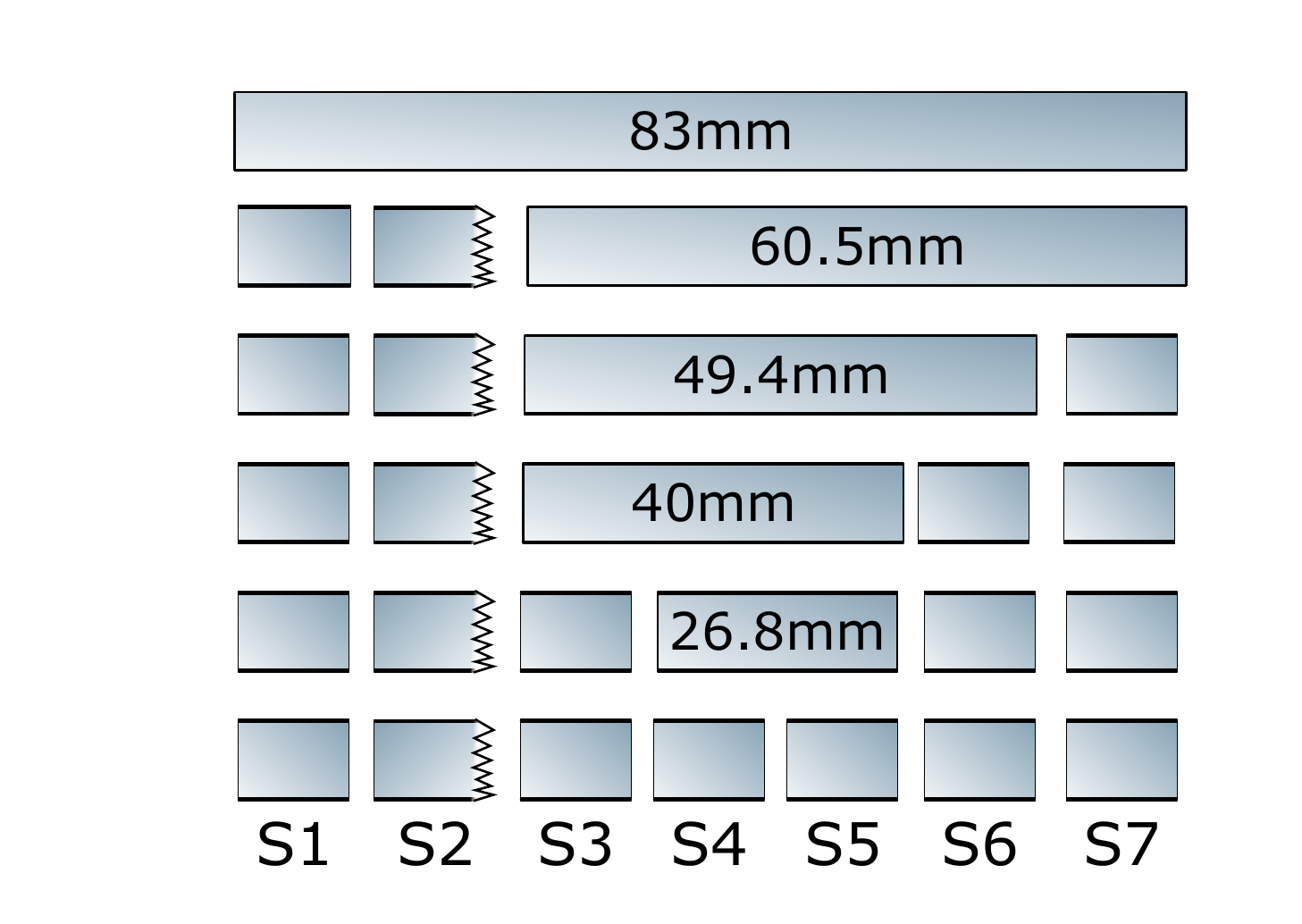}
\caption{Sequence of sample dicing. All the final sections are $\sim$10mm long. 
While polishing the 70mm-long sample, roughly 8mm of sample snapped from the left side and thus no phasematching spectrum could be recorded for S2 and for the 70mm-long sample.}
\label{img:chopping}
\end{figure}

\begin{figure}[bt]
\centering
\includegraphics[width = 0.7\textwidth]{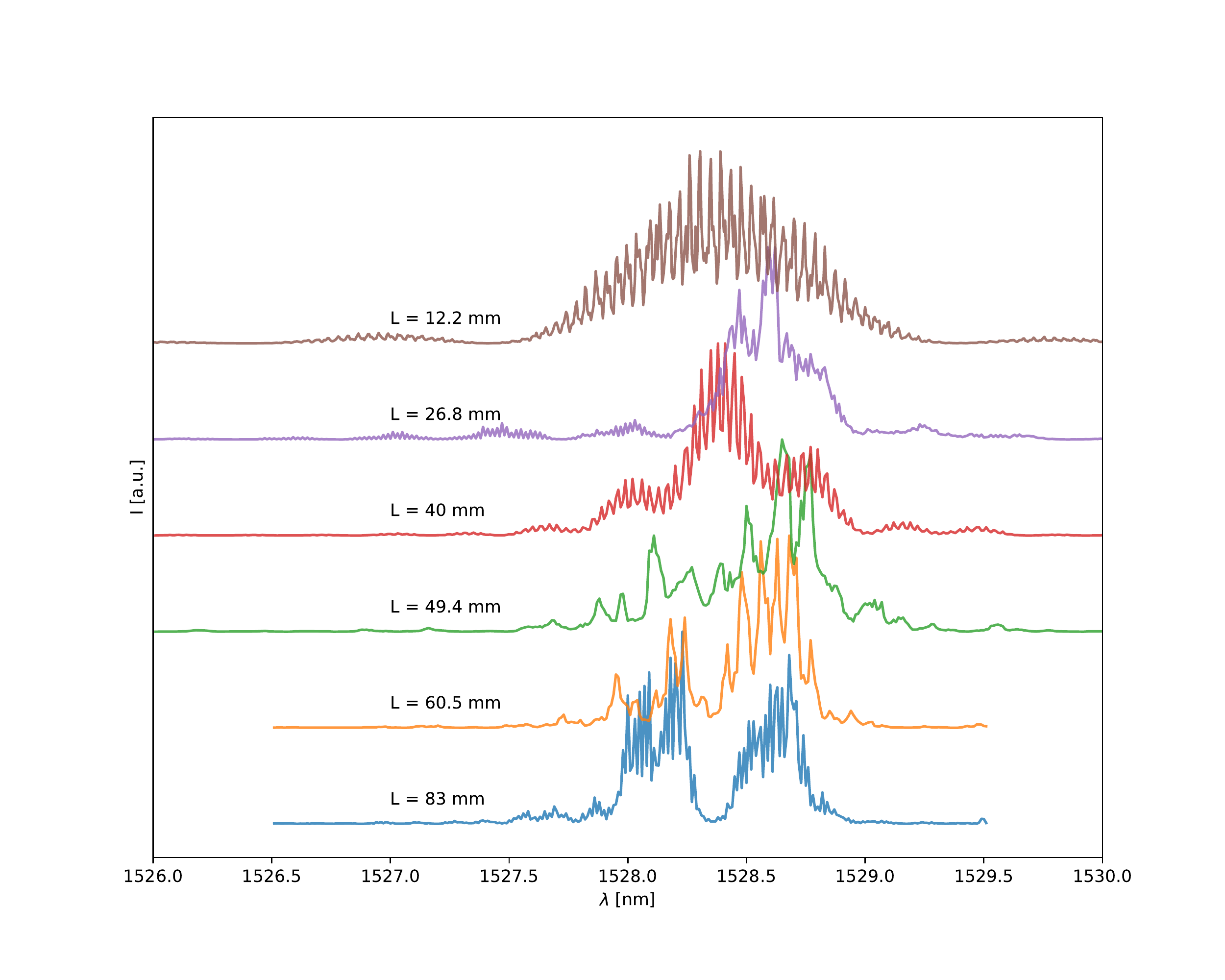}
\caption{Type 0 SHG phasematching spectra of a single waveguide for different waveguide lengths. Note that, as expected, the phasematching spectrum broadens as the sample becomes shorter and its shape tends towards the ideal sinc$^2$ profile. The oscillations are due to the presence of Fabry-Perot oscillations from the uncoated end facets.}
\label{img:ph_wg16}
\end{figure}

As expected, the spectra of the long waveguides are distorted due to the presence of waveguide inhomogeneities, as can be seen in Figure \ref{img:ph_wg16}.
A $\sim$1cm-long piece was then cut from one end of the original sample and the resulting surfaces were polished. Similar to the measurement for the original long sample, both resulting lengths were then temperature stabilised and the phasematching spectra of the waveguides in each section were again recorded. This process was repeated until the full sample was finally cut down into 7 pieces of approximately 1cm long, as shown in Figure \ref{img:chopping}. Note that some of the length is lost in the dicing and polishing of the sample as well as a small piece that was damaged during dicing.

Theoretical calculations show that, for a given inhomogeneity profile, the phasematching spectrum should become less distorted as the length of the waveguides is reduced, as waveguide imperfections become less critical \cite{Santandrea2019}. The phasematching spectra measured from a single waveguide as it was gradually cut into shorter and shorter pieces is shown in Figure \ref{img:ph_wg16}. As expected, one can see that the measured phasematching spectra gradually approach the ideal sinc$^2$-like shape as the sample is shortened down to 1cm. 

\begin{figure}[bt]
\centering
\includegraphics[width = 0.7\textwidth]{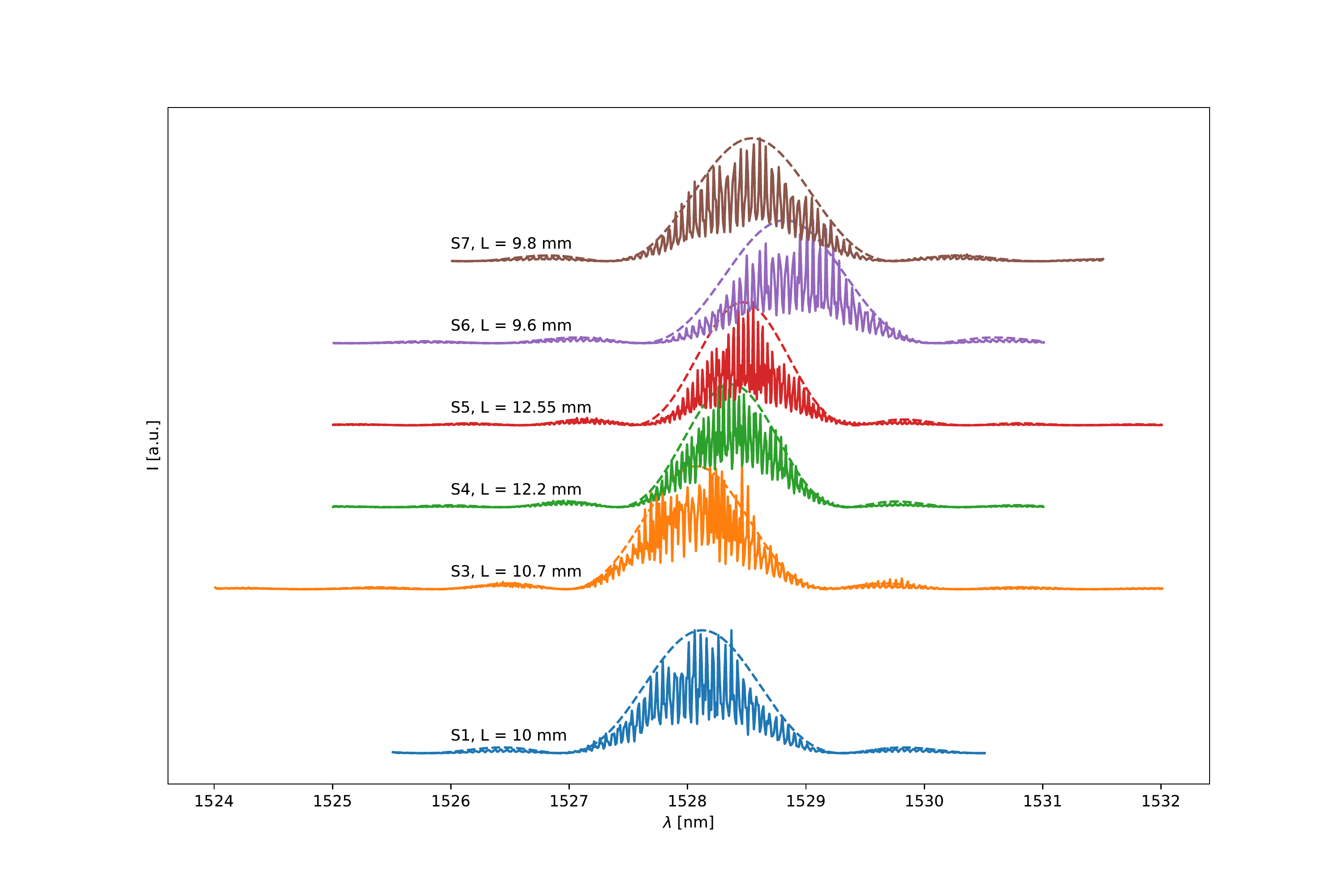}
\caption{Measured phasematching spectra for the different 1cm-long pieces in a single waveguide (WG1). The dashed lines correspond to the theoretical spectra fitted only for central wavelength. The shift of the phasematching centre is due to local variation of the waveguide properties and is used to derive the phasematching variation along the sample. Section S2 is missing since it broke during the dicing stage.}
\label{img:ph_wg16_1cm}
\end{figure}

From the measured phasematching profile of each 1cm-long piece a more quantitative investigation is undertaken. Figure \ref{img:ph_wg16_1cm} shows the measured phasematching profiles for all 1cm pieces across a single waveguide. For reference, the dashed curve shows the expected sinc$^2$ profile fitted only for the central wavelength. The central phasematching wavelength $\lambda_{pm}$ is found from the data as the average of the measured wavelengths weighted by the corresponding measured intensities. 
Since the sample is temperature stabilized, the deviation of $\lambda_{pm}$ from the target phasematching wavelength $\bar{\lambda}_{pm}=$1528.4nm is a measure of the waveguide inhomogeneities. With the help of  Eq. (\ref{eq:deltabeta}), one can quantify the phase mismatch $\delta\beta$ with respect to the ideal phasematching for each 1cm-long section.

\begin{figure}[hbt]
\centering
\includegraphics[width = 0.7\textwidth]{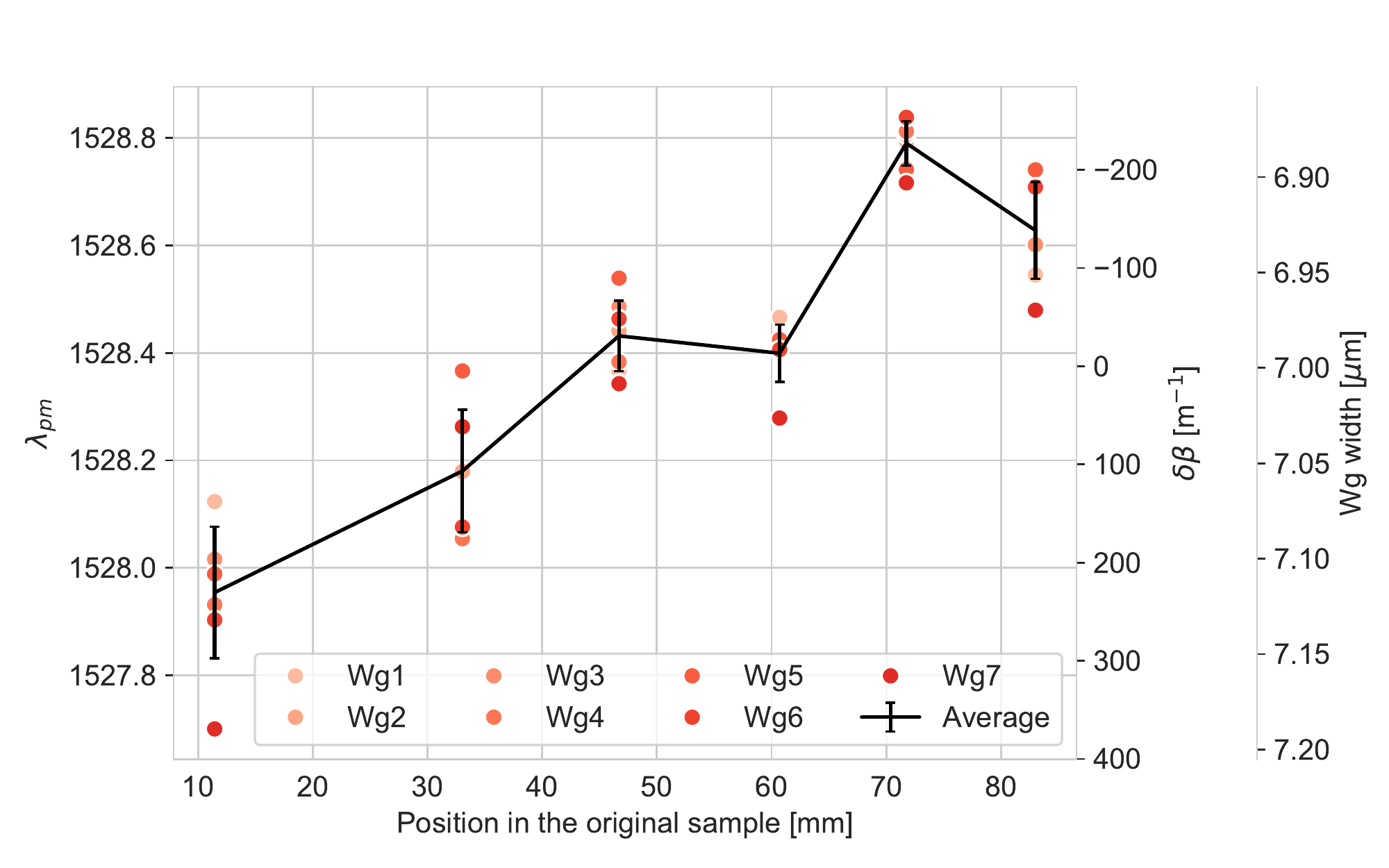}
\caption{Measured variation in the phasematching wavelength $\lambda_{pm}$ across all waveguides for each 1cm-long sample. The black solid line is the average $\lambda_{pm}$ shift. The corresponding phase mismatch variation $\delta\beta$ and waveguide width $w$ can be read on the right axes. Here, we assume that only the width variation is responsible of the whole phase mismatch $\delta\beta$ variation. }
\label{img:phasematching_shift}
\end{figure}

This characterisation is repeated for all remaining waveguides and the central phasematching wavelengths found in the 1cm samples for all waveguides are shown in Figure \ref{img:phasematching_shift}. 
It is immediately apparent that all the measured waveguides show a similar trend, whereby $\lambda_{pm}$ increases along the sample length.
Additionally, $\lambda_{pm}$ changes quite dramatically between the different sections, with an average maximum variation of 0.6nm. 
This corresponds to a maximum $\delta\beta$ variation of $\sim$442 m$^{-1}$.

Using the model presented in \cite{Strake1988}, one can relate the phasematching shift to a variation of the waveguide properties. 
For simplicity, we assume that only variations in the waveguide width contribute to phasematching variations along the sample, since this parameter is likely to be the major contributor to waveguide imperfections \cite{Santandrea2019}.
For the waveguides under consideration, a variation of $\delta\beta\sim$ 440m$^{-1}$ corresponds to a width inhomogeneity $\delta w\sim 0.25\mu$m, in close agreement with the theoretical estimation presented in \cite{Santandrea2019}. 
Interestingly, the retrieved $\delta w$ is close to the width error estimated in \cite{Chang2014} ($\delta w \sim 0.2\mu$m) with a completely different measurement technique, for a very different process realized in RPE-LN waveguides. The only similarity between the two systems is the employment of a wet-etching based photolithographic step for waveguide patterning. This is perhaps an indication that the main cause of waveguide imperfections resides in the photolithographic step.

\begin{figure}[hbt]
     \centering
	     \includegraphics[width = 0.8\textwidth]{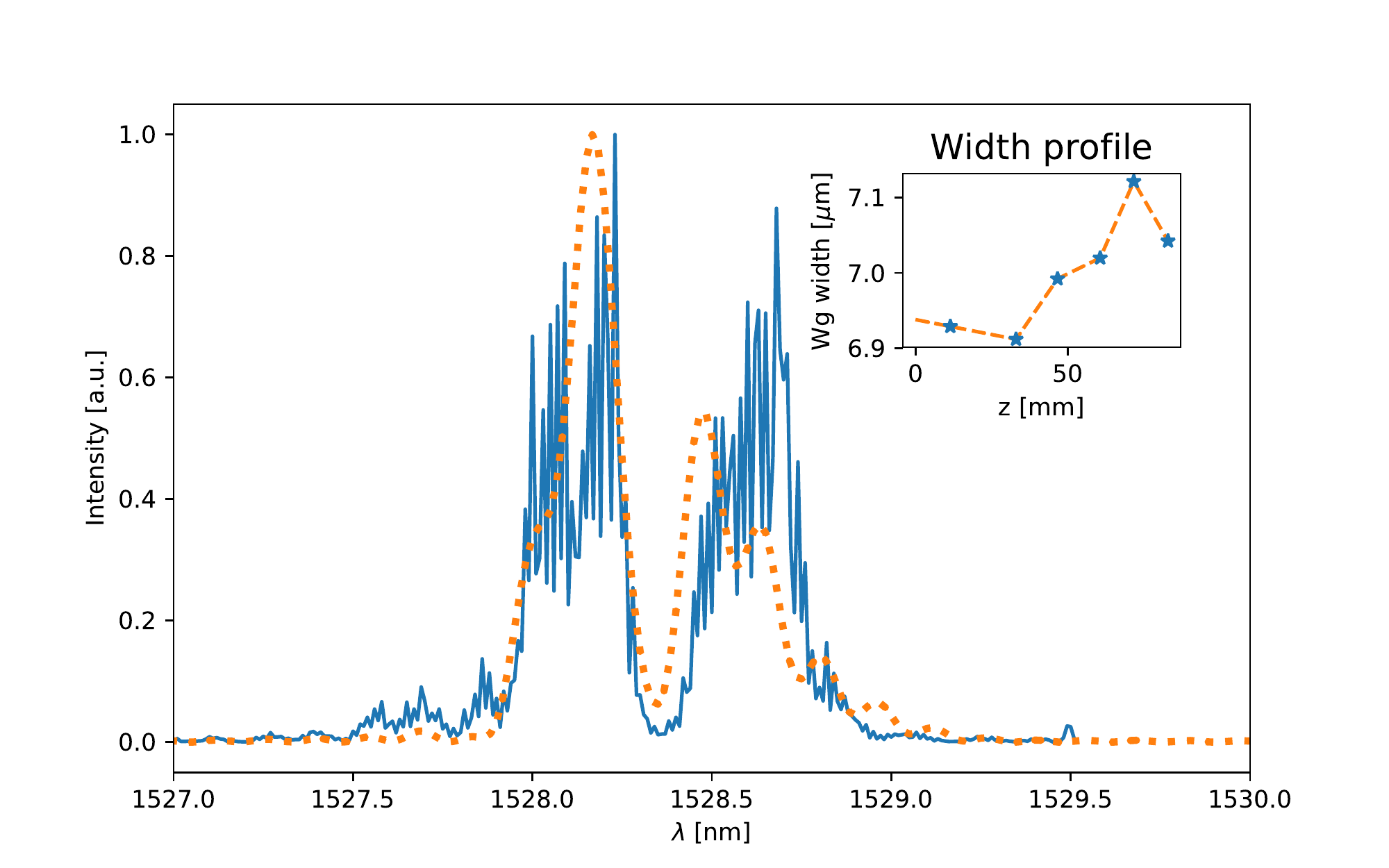}
     \caption{Reconstructed phasematching spectrum from the measured data points (solid blue line) and comparison with the measured one (dashed orange line). In the inset, the inferred waveguide widths and the linear interpolation used to reconstruct the phasematching are shown.}
     \label{img:pm_recon}
\end{figure}

Finally, it should be possible to reconstruct the original phasematching spectrum given the measured $\delta w$ profile along the sample. The performance of this reconstruction will of course be limited by the resolution of the measured $\delta w$ profile and hence only the profile of the full length sample is reconstructed. 
We interpolate the measured data points with a piecewise linear interpolation to avoid numerical artifacts that are seen to arise from the choice of higher order polynomials fits.
Thus, we obtain an approximate waveguide width profile $w(z)$ and the respective phase mismatch variation $\delta\beta(z)$. Eq. (\ref{eq:pm_integral}) is then used to calculate the expected phasematching spectrum of the original, full-length sample.
In approximately half of cases, the reconstruction of the phasematching was successful and showed very similar behavior to the phasematching that was initially measured from the full-length sample. One such reconstruction and the measured phasematching spectrum is shown in Figure \ref{img:pm_recon}. It is believed that the reconstruction failed in half of cases due to an insufficient spatial resolution for $\lambda_{pm}$. Note that, in contrast to previous work in PCF fibres \cite{FrancisJones2016}, no fitting has been performed to reproduce the measurement of the original sample. This shows that, in most cases, the phasematching properties in Ti:LN waveguide can be considered constant within 1cm.

\section{Conclusion}
In this paper, the inhomogeneity present in Ti:LN waveguides produced via UV photolithograpy was characterised by investigating the performance of a type 0 SHG process in these waveguides. The phasematching spectrum of seven waveguides was measured for the full-length sample (83mm) and for each resulting smaller section as $\sim$10mm-long pieces were cut from the ends of the initial sample.
The variation in the central phasematching wavelength along the sample was recorded for the 1cm pieces. From these measurements the variation in the phasematching across the sample were inferred, showing that the maximum phasematching deviation is $\delta\beta \leq$ 440m$^{-1}$. Under the assumption that variations in the phasematching are primarily due to imperfections in the width of the waveguides $\delta w$, a maximum width error of $\delta w\leq 0.25\mu$m was found, consistent with previous theoretical predictions \cite{Santandrea2019}.
It was also shown that the measured phasematching variation along the sample could be used to directly reconstruct the measured, highly distorted phasematching profile of the full-length sample. 
These results confirm the theoretical model presented in \cite{Santandrea2019}, thus verifying the validity of the conclusions presented therein. 

As the retrieved waveguide width error $\delta w$ is very close to values reported from other groups for different types of samples, we can thus conclude that this error is most likely the current ultimate resolution for waveguide fabrication using state-of-the-art UV masked photolithography. 
This provides crucial information that can be fed into the design of new samples. Knowing the fabrication limitations allows one to calculate the expected phasematching degradation for any desired process, particularly for high demand applications. Furthermore, understanding these limitations provides the foundation for new developments towards improved sample design and fabrication technologies.

\section{Acknowledgment}
The work was supported by the European Union via the EU quantum flagship project UNIQORN (Grant No. 820474) and by the DFG (Deutsche Forschungsgemeinschaft).

\end{document}